# Tractography with T1-weighted MRI and associated anatomical constraints on clinical quality diffusion MRI

Tian Yu[a], Yunhe Li[b], Michael E. Kim[a], Chenyu Gao[c], Qi Yang[a], Leon Y. Cai[b], Susane M. Resnick[f], Lori L. Beason-Held[f], Daniel C. Moyer[a], Kurt G. Schilling[d, e], Bennett A. Landman[a, b, c, d, e]

[a]Department of Computer Science, Vanderbilt University, Nashville, TN, USA; [b]Department of Biomedical Engineering, Vanderbilt University, Nashville, TN, USA; [c]Department of Electrical Engineering, Vanderbilt University, Nashville, TN, USA; [d]Department of Radiology and Radiological Sciences, Vanderbilt University, Nashville, TN, USA; [e]Vanderbilt University Institute of Imaging Science, Vanderbilt University, Nashville, TN, USA; [f]Laboratory of Behavioral Neuroscience, National Institute on Aging, Baltimore, MD, USA;

## ABSTRACT

Diffusion MRI (dMRI) streamline tractography, the gold standard for in vivo estimation of brain white matter (WM) pathways, has long been considered indicative of macroscopic relationships with WM microstructure. However, recent advances in tractography demonstrated that convolutional recurrent neural networks (CoRNN) trained with a teacher-student framework have the ability to learn and propagate streamlines directly from T1 and anatomical contexts. Training for this network has previously relied on high-resolution dMRI. In this paper, we generalize the training mechanism to traditional clinical resolution data, which allows generalizability across sensitive and susceptible study populations. We train CoRNN on a small subset of the Baltimore Longitudinal Study of Aging (BLSA), which better resembles clinical protocols. Then, we define a metric, termed the epsilon ball seeding method, to compare T1 tractography and traditional diffusion tractography at the streamline level. Under this metric, T1 tractography generated by CoRNN reproduces diffusion tractography with approximately two millimeters of error.

## 1. INTRODUCTION

Diffusion tractography is premised on identifying connections between parts of the brain through white matter pathways known as tracts[1]. These tracts are connected line segments that are often interpreted through multiple ways of association or connectivity[2]. Identifying proper termination criteria, shape characteristics, seeding criteria, etc. are all areas of active research. One area of recent controversial innovation[3] is in tractography generated from T1 weighted MRI, a tractography approach proposed by Cai et al.[4] based on convolutional recurrent neural networks (CoRNN) that creates streamlines without the use of diffusion information. Cai et al. compare bundles, or groups of streamlines sharing a similar trajectory or representing white matter pathways of the brain, in T1 weighted tractography and diffusion tractography and find that T1 tractography retains a lot of the same information contained in traditional diffusion tractography[5]. However, no streamline-to-streamline comparison between the two methods was performed.

Identifying which streamlines can be seen on T1 tractography and which cannot be seen is important. This allows us to measure equivalence or near equivalence between tractography methods. In this work, we propose epsilon ball seeding metric to do this by comparing individual streamlines. We account for early streamlines termination, unresolved directionality for streamlines, direction change at cross-fiber junctions, and occasional long outlier path streamlines. We explore how we model and capture the streamlines by applying the CoRNN T1 tractography method to a clinically feasible imaging cohort from an existing longitudinal study and retrain an algorithm appropriate for clinical resolution data. This refined algorithm is then applied to a sample of 9 individuals from the cohort, distinct from the training set. The study encompasses both traditional tractography on a scan-rescan basis and T1w tractography on the same subjects within the same scanning session.

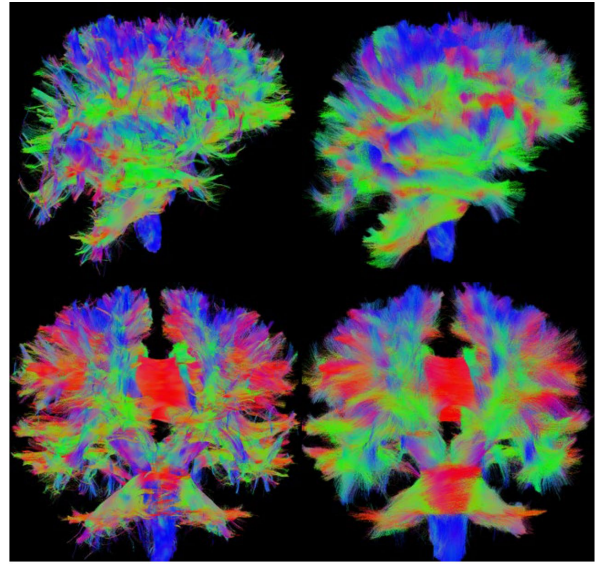

**Diffusion Tractography**     **T1 Tractography**

Figure 1. Tractography is the process of mapping out extended connections in the brain. Historically, tractography has been only done with dMRI data. Recent work has shown that similar structures can be learned from only T1 images and the anatomical context they provide. Visually the connections are incredibly similar. No prior work has explored the difference on a streamline-by-streamline basis.

## 2. METHODS

We choose a set of 15 subjects from the Baltimore Longitudinal study of Aging (BLSA). All subjects were screened for artifacts and significant distortion problems, but otherwise are selected randomly. Each of the subjects has a pair of scan and rescan diffusion MRI as well as an associated T1 weighted MRI. 3T Phillips scanner were used to acquire dMRI in 32 directions at b value of 700 s/mm$^2$. Initial voxel dimension of 2.2 x 2.2 x 2.2 mm$^3$ were resampled to 0.8125 x 0.8125 x 0.8125 mm$^3$. T1 weighted MRI were acquired on the same scanner at voxel dimension of 1.2 x 1 x 1 mm$^3$, resampled to 1.0 x 1.0 x 1.0 mm$^3$.

We process these data using PreQual which includes denoising, susceptibility induced distortion correction, eddy current induced distortion correction and quality assurance[6]. The dataset is partitioned into a training set of 5 subjects (8 sessions), a validation set with 1 subject (3 sessions), and a testing set of 9 subjects. Notably, in the testing set, although subjects may have undergone multiple scans, we randomly include on session per subject to maintain consistency.

### 2.1 Data Preprocessing

Following previous work on T1 tractography[4], we perform the same preprocessing steps to prepare data for the CoRNN network. For T1 weighted MRI, we first compute brain mask and perform N4 bias correction[7,8]. Utilized MRtrix3's five-tissue-type segmentation algorithm[9], we segment T1 weighted MRI into grey matter, subcortical grey matter, white matter, cerebrospinal fluid, and pathological tissue. The resulting images contains valuable anatomical context. In addition, SLANT deep learning framework was used to segment 132 brain regions, and then merged into 46 larger regions defined by the BrainColor protocol[10]. Finally, we use WM learning (WML) framework to compute 72 WM bundle regions defined by the TractSeg algorithm[11,12]. All anatomical information is one-hot encoded for a total of 123 input channels. We first align the bias-corrected T1 images to the Montreal Neurological Institute (MNI) atlas space with 1 mm isotropic voxel resolution. Subsequently, the T1 images, brain mask, and associated anatomical context are transformed to the MNI common space with a 2mm isotropic resolution, leveraging the transformation matrix from the previous registration.

To prepare the dMRI, we first compute average b0 and rigidly register the result to T1w MRI space. Subsequently, we estimate fiber orientation distribution function (FODs) using constrained spherical deconvolution (CSD)[13]. We then generate 1 million streamlines using MRTrix3 SDStream's anatomically constrained deterministic tractography with seeds set at the GM/WM boundary. We use a 1mm step size. The minimum/maximum length of streamline is 50/250mm. Due to GPU memory constraints, we shuffle the 1 million streamline and split into batches of 1000 streamlines. Since we convert each step between adjacent point on a streamline into unit vectors in spherical coordinates, the resulting labels are one point shorter than the length of streamlines. To match them, we drop the last point of all streamlines. Tractography, originally in b0 space, is rigidly transformed to T1 space and, then, to 2mm MNI space.

### 2.2 Network architecture and training paradigm

Following our previous work in T1 tractography on Human Connectome Project[4], we train a CoRNN model to produce streamlines. The teacher-student framework was implemented, wherein the teacher model utilized FODs as inputs, and the student model incorporated T1 and anatomical context as inputs. For the teacher model, we perform trilinear interpolation at streamline point locations and use four layers of MLP blocks to create an embedding of FOD information. This operation provides the network spatial information for performing streamline propagation. Each MLP block consists of a linear layer of size 512, a batch normalization layer with no running estimate of mean and variance during evaluation and leaky ReLU activation function with a negative slope of 0.1. The embedding is then passed to two stacked gated-recurrent-unit (GRU) with hidden size of 512 to encode streamline memory. The output of MLP and GRU are concatenated together, yielding an intermediate vector of size 1024, which is then decoded with a linear layer to the output vector of size 2. The output vector represents azimuth and zenith in spherical coordinates. Due to GPU memory constraints, we use batch size of 1 on the image level. For each subject, randomly selected batch of 1000 streamlines are passed for the trilinear interpolation. The loss function is cosine similarity loss to enforce the output of the network to be close to the labels.

For the student model, we used frozen weights from the best epoch of the teacher model for the GRU layer and decoding linear layer. The main task here is to learn similar embedding of FOD information from T1 and anatomical context. A single 7 x 7 x 7 3D convolutional project layer is used to extract information. This layer provides 1.4 x 1.4 x 1.4 cm receptive field in 2mm space, which is enough for identification of grey matter and white matter boundary according to Cai et al[4]. Besides this change, the student model has the same architecture as the teacher model: trilinear interpolation at streamline points, four block MLP layers, GRU layers and one decoding linear layer. In additional to cosine similarity loss against the label, we also add a contrastive loss between the output of MLP of the teacher model and the output of MLPs of the student model.

We use an Adam optimizer with initial learning rate of 0.001. We stop training when there is no improvement in validation loss after 200 epochs. The model weights with lowest validation loss are used for inference and evaluation.

### 2.3 Inference and stopping criterion

The student network is used during inference. Propagation of streamlines begin with trilinear interpolation to probe the image grid at the seed location. Based on previous work, we randomly seed the starting points within the white matter, specifically choosing points where the five-tissue-type segmented tissue mask indicates a non-zero value[4]. The hidden state of GRU is initialized to zero. From there, the model predicts the next point of the streamlines. The tracking process continues until it meets the stopping criteria as defined in previous research.[4] Unlike GM/WM interface seeding, seeds might appear in the middle of the brain. Consequently, we adjust our method by flipping the streamlines for tracking in the reverse direction. One caveat is that since we initialize the hidden state of GRU to zero, the hidden state is not useful at first. Thus, we discard the first five predicted points when streamlines are terminated in one direction.

Unlike traditional tractography, the tracking process occurs on the step level. To accelerate the tracking process, we implement tracking in a batch. For each batch, every streamline either steps to the next point, terminates, or is rejected. Once bi-directional tracking completes, a new batch of streamlines is seeded. The tracking ends when there are more than or equal to 1 million streamlines generated.

### 2.4 Epsilon Ball seeding method

In prior work, a symmetric distance function named minimum average direct flip (MDF) distance was introduced as a means to quantify the similarity between two streamlines[14]. This method is applicable only when the streamlines have an equivalent number of points. In our study, we operated under the premise that each streamline being evaluated would be resampled to have the same number of points. Accordingly, we resampled every streamline to consist of 100 points using linear interpolation. Expanding on the concept of MDF distance, we introduce a non-symmetric method for comparing individual streamlines from two registered tractograms, which we denote as tractogram A and tractogram B. For each streamline in tractogram A, we establish an epsilon ball, centered at its seed point, with a 0.5 voxel radius. Streamlines in tractogram B that intersect this epsilon ball are marked as matching the streamline from tractogram A. We then compute the MDF distance for the selected streamlines in tractogram A against all matched streamlines from tractogram B. The streamlines in tractogram B with the smallest MDF distance is designated as the optimal match. Using this method, we allocate a distance value to each streamline in A. This informs us about the proximity of a streamline in A to its most compatible match in tractogram B. Note that there is possibility that no streamline in tractogram B intersects the defined epsilon ball, and we assign a value of infinity to the such streamlines in A.

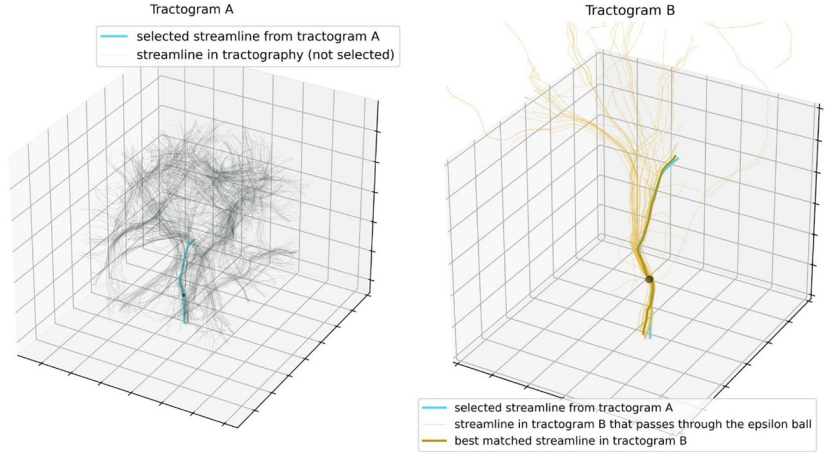

Figure 2. Illustration of the Epsilon Seeding Method. On the left, Tractogram A displays 1,000 random streamlines, with one highlighted to indicate selection. On the right, Tractogram B shows streamlines intersecting the epsilon ball of the selected streamline from Tractogram A. The best-matched streamline is highlighted in blue. Additionally, the selected streamline from Tractogram A is superimposed on Tractogram B for direct comparison.

## 3. RESULTS

For each subject in the testing set, three tractograms were derived from a pair of dMRI scans and a T1w MRI scan: (1) A tractogram produced using MRtrix3 SDStream tractography algorithm applied to the first dMRI scan (dMRI 1), (2) A tractography generated using the same SDStream tractography algorithm but applied to the second dMRI scan (dMRI 2), (3) A tractogram based on our method applied to the T1w MRI scan.

To evaluate the tractogram generated by our method, the epsilon seeding metric was employed on three distinct pairs: (a) A comparison between the tractogram generated by SDStream algorithm from dMRI 1 (tractogram A) and dMRI 2 (tractogram B), (b) A comparison between our T1 tractography method (tractogram A) and SDStream algorithm generated tractogram from dMRI 1 (tractogram B), (c) A comparison between tractogram generated by SDStream algorithm on dMRI 1 (tractogram A) and tractogram produced using our method on T1w MRI (tractogram B). To reduce computation complexity, we sample 50,000 streamlines from each tractography for the purpose of comparison. The comparison of scan-rescan variability serves as a reference for the inherent variability of tractography, which helps contextualizing the precision of our results. Of note, this comparison is not intended to equate the sources of variance in scan-rescan and the source of variance in our method.

Table 1. Key statistics on three comparing pairs using epsilon seeding metrics.

| Comparison pairs | Mean of Means (mm) | Mean of Medians (mm) | Average outlier streamlines (%) |
| --- | --- | --- | --- |
| Scan to Rescan | 4.22016 | 3.15576 | 0.0973 |
| T1 to diffusion | 6.09626 | 4.75834 | 0.134 |
| Diffusion to T1 | 5.76262 | 4.44779 | 0.10066 |

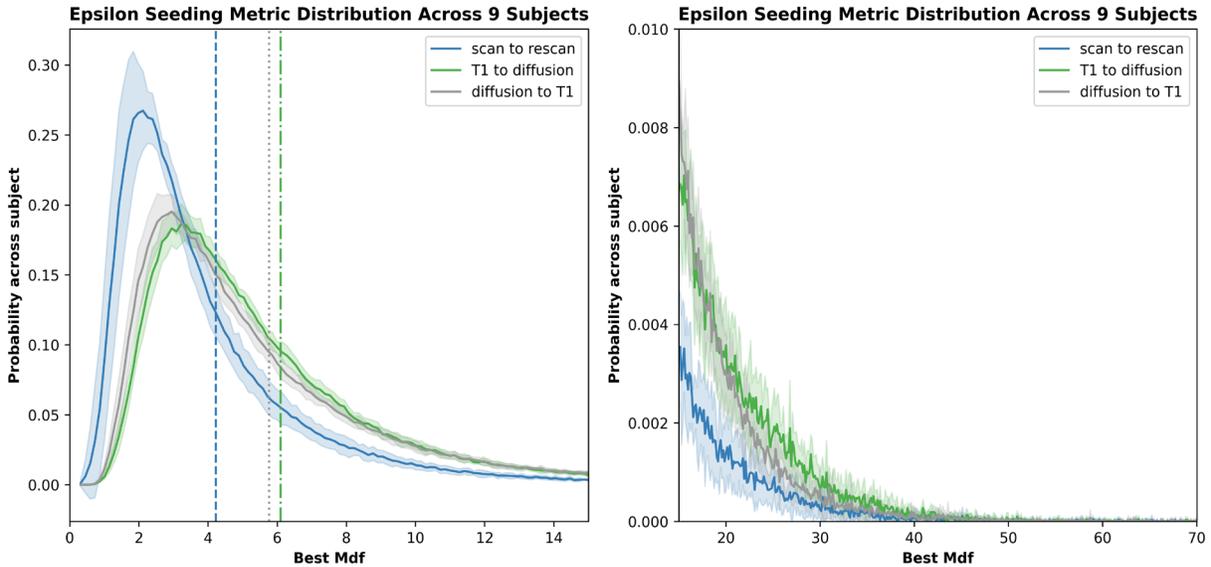

Figure 3. Epsilon seeding metrics on three pairs of tractograms. The mean and standard deviation of epsilon seeding metric across 9 subjects are plotted. The dash vertical lines are the median.

We find that T1 tractography reproduces the SDStream tractography on diffusion data with two millimeters of additional error on average. For each streamline in diffusion tractography, we can find a streamline with 1.29 millimeters of additional error based on the mean of medians, indicating high sensitivity in T1 tractography to diffusion tractography. However, the specificity of the tracts identified with T1 tractography is slightly worse. For each streamline in T1 tractogram, there exists a streamline with 1.6 millimeter of additional error based on mean of medians compared to scan rescan variability.

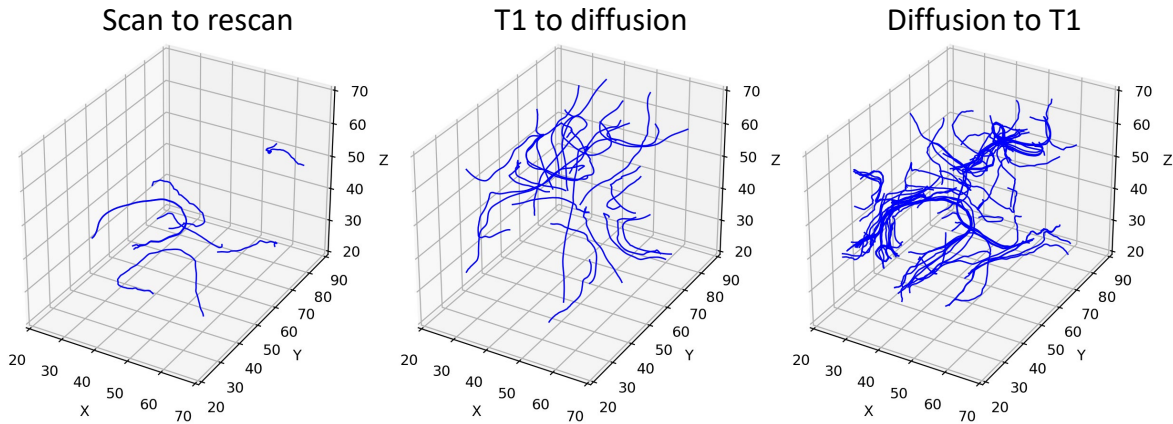

Figure 4. Erroneous streamlines of random subject from the testing set are presented. These streamlines are outliers because they cannot capture any streamlines in the epsilon ball.

Scan to rescan outliers (Fig. 4) appear to be sparse and have a few distinct trajectories. The additional tracts in T1 over to diffusion is denser and appear to be cross hemisphere or longitudinal. It has more branching at potential choke points or cross fiber regions without specific a tract structure. Finally, the additional tracts that are captured by diffusion tractography but not in T1 tractography appear to be associated with anterior and posterior commissure tracts as indicated by the U fibers. These tracts are known to be difficult to associate in noisy data or in low quality data[15]. Further investigation into tracts that are found in diffusion tractography but not in T1 tractography is likely needed.

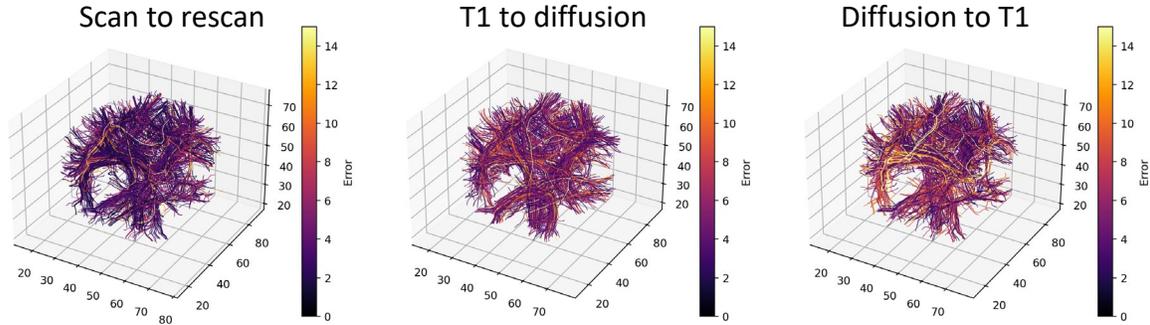

Figure 5. Visualization of 1000 random streamlines colored by their best MDF of a random subject from the testing. (Not the same subject as figure 4)

From Fig. 5, we can see that diffusion scan to rescan error under our epsilon ball metrics is largely spatially consistent. We do not see much anatomical variation. Comparing streamlines in T1 relative to diffusion tractography, we see a larger error (table 1). The difference is largely correlated in streamlines that are cortically adjacent. This is consistent with Fig. 4, which shows the T1 to diffusion streamline outliers are largely observed in cortical association areas. When we look at the differences in diffusion relative to T1, we see the systematic differences in anterior and posterior cross fiber regions.

## 4. DISCUSSION

In this work, we demonstrate that convolution recurrent network can converge on not only high angular diffusion MRI but also clinical quality diffusion MRI. To address the need to better understand the difference between diffusion tractography and T1 tractography, we propose the epsilon ball seeding metric. Prior work proposed other streamline level comparisons[16–19]. We demonstrate that a model trained on only 6 subjects is able to reproduce diffusion tractography with approximately 2 mm of additional error compared with scan-rescan. These results are very surprising because T1w MRI is believed to lack microstructural information on which tractography algorithms depend. While demonstrating great promise, the preliminary results show that T1 tractography has less optimal performance on anterior posterior commissure tracts and tracts that are cortically adjacent. Further research is needed to determine if these tracts are learnable, and to characterize the generalizability of the emerging T1 tractography findings.

## 5. ACKNOWLEDGEMENT

This work was conducted in part using the resources of the Advanced Computing Center for Research and Education at Vanderbilt University, Nashville, TN. This work was supported by the National Institutes of Health (NIH) under award numbers 5R01EB017230, 1U34DK123895-01, U34DK123894-01, P50HD103537, U54HD083211, U54HD083211-S1, K01EB032898, R01NS095291, and T32GM007347 and by the National Science Foundation (NSF) under award number 2040462. This research was conducted with the support from the Intramural Research Program of the National Institute on Aging of the NIH. The Vanderbilt Institute for Clinical and Translational Research (VICTR) is funded by the National Center for Advancing Translational Sciences (NCATS) Clinical Translational Science Award (CTSA) Program, Award Number 5UL1TR002243-03. The content is solely the responsibility of the authors and does not necessarily represent the official views of the NIH or NSF.

## REFERENCES

[1] Assaf, Y. and Pasternak, O., "Diffusion tensor imaging (DTI)-based white matter mapping in brain research: a review," J Mol Neurosci **34**(1), 51–61 (2008).

[2] Lee, S. K., Dong, I. K., Kim, J., Dong, J. K., Heung, D. K., Dong, S. K. and Mori, S., "Diffusion-tensor MR imaging and fiber tractography: a new method of describing aberrant fiber connections in developmental CNS anomalies," Radiographics **25**(1), 53–65 (2005).

[3] "Convolutional-recurrent neural networks approximate diffusion tractography from T1-weighted MRI and associated anatomical context | OpenReview.", <https://openreview.net/forum?id=TelM2TBYgoA> (1 August 2023 ).
[4] Cai, L. Y., Lee, H. H., Newlin, N. R., Kerley, C. I., Kanakaraj, P., Yang, Q., Johnson, G. W., Moyer, D., Schilling, K. G., Rheault, F. and Landman, B. A., "Convolutional-recurrent neural networks approximate diffusion tractography from T1-weighted MRI and associated anatomical context," bioRxiv (2023).
[5] Jeurissen, B., Descoteaux, M., Mori, S. and Leemans, A., "Diffusion MRI fiber tractography of the brain," NMR Biomed **32**(4) (2019).
[6] Cai, L. Y., Yang, Q., Hansen, C. B., Nath, V., Ramadass, K., Johnson, G. W., Conrad, B. N., Boyd, B. D., Begnoche, J. P., Beason-Held, L. L., Shafer, A. T., Resnick, S. M., Taylor, W. D., Price, G. R., Morgan, V. L., Rogers, B. P., Schilling, K. G. and Landman, B. A., "PreQual: An automated pipeline for integrated preprocessing and quality assurance of diffusion weighted MRI images," Magn Reson Med **86**(1), 456–470 (2021).
[7] Fischl, B., "FreeSurfer," Neuroimage **62**(2), 774–781 (2012).
[8] Tustison, N. J., Avants, B. B., Cook, P. A., Zheng, Y., Egan, A., Yushkevich, P. A. and Gee, J. C., "N4ITK: Improved N3 Bias Correction," IEEE Trans Med Imaging **29**(6), 1310 (2010).
[9] Tournier, J. D., Calamante, F. and Connelly, A., "MRtrix: Diffusion tractography in crossing fiber regions," Int J Imaging Syst Technol **22**(1), 53–66 (2012).
[10] Huo, Y., Xu, Z., Xiong, Y., Aboud, K., Parvathaneni, P., Bao, S., Bermudez, C., Resnick, S. M., Cutting, L. E. and Landman, B. A., "3D whole brain segmentation using spatially localized atlas network tiles," Neuroimage **194**, 105–119 (2019).
[11] Yang, Q., Hansen, C. B., Cai, L. Y., Rheault, F., Lee, H. H., Bao, S., Chandio, B. Q., Williams, O., Resnick, S. M., Garyfallidis, E., Anderson, A. W., Descoteaux, M., Schilling, K. G. and Landman, B. A., "Learning white matter subject-specific segmentation from structural MRI," Med Phys **49**(4), 2502–2513 (2022).
[12] Wasserthal, J., Neher, P. and Maier-Hein, K. H., "TractSeg - Fast and accurate white matter tract segmentation," Neuroimage **183**, 239–253 (2018).
[13] Tournier, J. D., Calamante, F. and Connelly, A., "Robust determination of the fibre orientation distribution in diffusion MRI: Non-negativity constrained super-resolved spherical deconvolution," Neuroimage **35**(4), 1459–1472 (2007).
[14] Garyfallidis, E., Brett, M., Correia, M. M., Williams, G. B. and Nimmo-Smith, I., "QuickBundles, a Method for Tractography Simplification," Front Neurosci **6**, 33886 (2012).
[15] Guevara, M., Guevara, P., Román, C. and Mangin, J. F., "Superficial white matter: A review on the dMRI analysis methods and applications," Neuroimage **212**, 116673 (2020).
[16] Calamante, F., "The Seven Deadly Sins of Measuring Brain Structural Connectivity Using Diffusion MRI Streamlines Fibre-Tracking," Diagnostics **9**(3) (2019).
[17] Soares, J. M., Marques, P., Alves, V. and Sousa, N., "A hitchhiker's guide to diffusion tensor imaging," Front Neurosci **7**(7 MAR) (2013).
[18] Mukherjee, P., Chung, S. W., Berman, J. I., Hess, C. P. and Henry, R. G., "Diffusion Tensor MR Imaging and Fiber Tractography: Technical Considerations," American Journal of Neuroradiology **29**(5), 843–852 (2008).
[19] Schilling, K. G., Nath, V., Hansen, C., Parvathaneni, P., Blaber, J., Gao, Y., Neher, P., Aydogan, D. B., Shi, Y., Ocampo-Pineda, M., Schiavi, S., Daducci, A., Girard, G., Barakovic, M., Rafael-Patino, J., Romascano, D., Rensonnet, G., Pizzolato, M., Bates, A., et al., "Limits to anatomical accuracy of diffusion tractography using modern approaches," Neuroimage **185**, 1 (2019).